\documentclass[journal]{IEEEtran}
\ifCLASSINFOpdf
  \usepackage[pdftex]{graphicx}
\else
  \usepackage[dvips]{graphicx}
\fi
\usepackage{amsmath}
\usepackage{algorithmic}
\ifCLASSOPTIONcompsoc
 \usepackage[caption=false,font=normalsize,labelfont=sf,textfont=sf]{subfig}
\else
 \usepackage[caption=false,font=footnotesize]{subfig}

\usepackage{stfloats}
\usepackage{url}

\usepackage{amssymb}
\usepackage{algorithm}

\usepackage{hyperref}
\usepackage{color}

\begin{document}
%

\title{ Unsupervised  Music  Source Separation Using Differentiable Parametric Source Models}

%
%
%
%
%
%

\author{Kilian~Schulze-Forster, Ga\"el Richard~\IEEEmembership{Fellow,~IEEE}, Liam Kelley, Clement S. J. Doire, and Roland Badeau~\IEEEmembership{Senior Member,~IEEE}
\thanks{Manuscript received July 26, 2021; revised January 21, 2022 and January~27, 2023. This project has received funding from the European Union (Horizon 2020 research and innovation programme - Marie Sk\l odowsa-Curie No. 765068 and ERC, HI-Audio, 101052978). Views and opinions expressed are however those of the author(s) only and do not necessarily reflect those of the European Union or the European Research Council. Neither the European Union nor the granting authority can be held responsible for them.}
\thanks{K. Schulze-Forster, G. Richard, and R. Badeau are with the Laboratoire de Traitement et Communication de l’Information (LTCI), T\'el\'ecom Paris, Institut Polytechnique de Paris, 91120 Palaiseau, France. (e-mail: \{kschulze, gael.richard, roland.badeau\}@telecom-paris.fr). C. S. J. Doire is with Sonos Inc., 75002 Paris, France. (e-mail: clement.doire@sonos.com)}
}

%
%

\markboth{Preprint - submitted for publication in IEEE/ACM Trans. on Audio, Speech, and Language Processing}%
{Shell \MakeLowercase{\textit{et al.}}: Bare Demo of IEEEtran.cls for IEEE Journals}
%



\maketitle

\begin{abstract}



Supervised deep learning approaches to underdetermined audio source separation achieve state-of-the-art performance but require a dataset of mixtures along with their corresponding isolated source signals. Such datasets can be extremely costly to obtain for musical mixtures. This raises a need for unsupervised methods. We propose a novel unsupervised model-based deep learning approach to musical source separation. Each source is modelled with a differentiable parametric source-filter model. A neural network is trained to reconstruct the observed mixture as a sum of the sources by estimating the source models' parameters given their fundamental frequencies. At test time, soft masks are obtained from the synthesized source signals. The experimental evaluation on a vocal ensemble separation task shows that the proposed method outperforms learning-free methods based on nonnegative matrix factorization and a supervised deep learning baseline. Integrating domain knowledge in the form of source models into a data-driven method leads to high data efficiency: the proposed approach achieves good separation quality even when trained on less than three minutes of audio. This work makes powerful deep learning based separation usable in scenarios where training data with ground truth is expensive or nonexistent.
\end{abstract}


%
\IEEEpeerreviewmaketitle

\section{Introduction}
\label{sec:intro}
%
%
%
%

\IEEEPARstart{A}{udio} source separation is the task of estimating the individual signals of several sound sources when only their mixture can be observed. When the sources are musical instruments (including singing voice), we refer to the task as Musical Source Separation (MSS) \cite{cano2018musical}. It has many applications, for example in up-mixing or re-mixing of recordings whose individual source signals are not accessible. It is also used to create play-along tracks for students of musical instruments. Furthermore, MSS is an important pre-processing step for several music information retrieval tasks such as automatic lyrics transcription \cite{demirel2020automatic}. 

Music mixtures are especially challenging because the source signals are usually highly correlated in time and frequency as opposed to speech or speech-noise mixtures \cite{vincent2018audio}. Beyond, certain instruments may be present multiple times as distinct sources in music mixtures, e.g. several singers in a choir. Hereafter, we refer to this issue as \textit{homogeneous sources}.

State-of-the-art performance in MSS is achieved by Deep Neural Networks (DNNs) which are trained in a supervised fashion \cite{takahashi2020d3net, defossez2019demucs, stoter2019open}. However, they have two shortcomings which we address in this paper.

Firstly, they are not able to separate homogeneous sources. For example, the methods in \cite{takahashi2020d3net, defossez2019demucs, stoter2019open} are able to separate \textit{all} singing voices from an instrumental accompaniment but provide only the mixture of these voices instead of further separating them into the different singer signals. Hence, they can neither be used to obtain only the \textit{lead} vocals nor to separate vocal ensembles or violin quartets, for example.

Secondly, they require training data with available ground truth, i.e. mixtures for which target source signals are available in isolation. However, such isolated signals are difficult, sometimes impossible, to obtain for music mixtures. If the instruments were recorded separately, the ground truth signals exist but are usually not distributed. This is usually the case for pop music. For most other genres such as jazz, classical music, or folk, it is common practice that the musicians perform together in the same room and only the mixture of the instrument signals is recorded. Hence, no isolated signal recordings exist. Special recording sessions may be arranged in order to record signals in isolation \cite{cuesta2018analysis}, however, this is not only extremely costly but also leads to unnatural conditions for the musicians.


Therefore, there is a need for separation methods that do not require ground truth signals for training. Such methods may be learning-free or unsupervised.

\textit{Learning-free} methods estimate all parameters directly from the test mixture \cite{vincent2018audio}. Hence, they do not require any training data. Nonnegative Matrix Factorization (NMF) \cite{lee1999learning} and its numerous extensions have successfully been used for learning-free MSS \cite{vincent2018audio}. Using side information such as musical scores \cite{ewert2012using, hennequin2011score} or fundamental frequency (F0) \cite{durrieu2011musically}, NMF-based methods can separate homogeneous sources.

\textit{Unsupervised} methods have a training stage and require only mixtures (no isolated sources) for learning. At test time, their parameters are fixed. They have the potential to provide superior performance similar to supervised methods while being less demanding regarding data. Recently proposed unsupervised deep learning methods for audio source separation are based on assumptions such that the sources are uncorrelated \cite{wisdom2020unsupervised, wisdom2021sparse} or not homogeneous \cite{narayanaswamy2020unsupervised, seetharaman2020bootstrapping}. Therefore, they are not applicable to music mixtures where sources are correlated and possibly homogeneous.



In this work, we propose and evaluate a novel approach to unsupervised source separation which does not make such assumptions. It is hence also applicable but not limited to music mixtures. The approach is inspired by the recent line of research which integrates signal processing models in DNNs to incorporate domain knowledge \cite{shlezinger2020model, engel2019ddsp}. Each source is modeled with a differentiable parametric source model. During training, the task of the DNN is to re-synthesize the observed mixture as a sum of the sources by estimating the source parameters. Separation is achieved because the F0s for all sources are estimated from the mixture and assigned to the sources beforehand. This can be done using existing methods such as \cite{cuesta2020multiple, schramm2017multi}. 

Besides being unsupervised and able to separate homogeneous sources, the approach has further advantages: high data efficiency as well as parametric, hence interpretable and modifiable, source estimates. 
Briefly, the contributions of this work are:
\begin{itemize}
    \item a novel unsupervised deep learning approach for audio source separation,
    \item the integration of parametric source models in deep learning based audio source separation,
    \item a new differentiable procedure to estimate \textit{stable} time-varying all-pole filters with a DNN using line spectral frequency parameterization,
    \item an extensive experimental evaluation of the proposed method on a musical source separation task and comparison to learning-free and supervised baselines,
    \item the open source code\footnote{\url{https://github.com/schufo/umss}} for the proposed method and experiments.
\end{itemize}
The rest of the paper is structured as follows: In Section~\ref{sec:related_work} we review related work on audio source separation and model-based deep learning. The proposed method is explained in Section~\ref{sec:method} and its experimental evaluation is outlined in Section~\ref{sec:experiments}. We present and discuss results in Section~\ref{sec:results_discussion} and conclude in Section~\ref{sec:conclusion}.

\section{Related Work}
\label{sec:related_work}

In this section we review work on homogeneous musical source separation, learning-free and unsupervised source separation, and, finally, on the integration of signal processing models in deep neural networks.

Homogeneous audio sources are not easily distinguishable in the time-frequency domain and pose a permutation problem \cite{hershey2016deep, yu2017permutation}. While permutation-invariant training is used for supervised speech separation \cite{yu2017permutation, kolbaek2017multitalker}, methods for musical homogeneous source separation exploit side-information such as F0 estimates \cite{petermann2020deep, durrieu2011musically} or a musical score \cite{gover2020score, ewert2012using, hennequin2011score} to guide the separation.  

Two deep learning approaches for supervised choir separation were proposed recently. In this context, a choir is composed of four homogeneous sources: a soprano, alto, tenor, and a bass singer. Petermann \textit{et al.} \cite{petermann2020deep} modified the conditioned U-Net \cite{meseguer2019conditioned} so that the target source can be selected and separated using its F0 information. Results show that this leads to improved objective separation quality compared to using non-informed source-specific models. However, ground truth source signals are needed for training and they are rare for choir recordings. This motivated Gover and Depalle \cite{gover2020score} to synthesize choir singing from MIDI files and to use this synthetic data for training of a score-informed DNN. When tested on real choir recordings, the model is outperformed by the learning-free, score-informed NMF proposed in \cite{ewert2012using}. This shows that the performance of supervised DNNs depends strongly on the quality and quantity of the training data.

Therefore, learning-free methods are a powerful alternative in limited data settings. Several separation methods based on NMF are learning-free and can exploit side-information to separate homogeneous sources. NMF approximates a spectrogram with a matrix product of two low-rank matrices containing spectral templates and their activations, respectively \cite{vincent2018audio}.
Ewert and Müller \cite{ewert2012using} proposed to initialize both templates and activations using musical score information. This leads to improvements compared to random initialization. Using the score allows even to separate notes played by the left and the right hand in piano recordings. Similarly, Hennequin \textit{et al.} \cite{hennequin2011score} used a musical score to initialize the activations whereas the templates consist of parametric frequency atoms. Durrieu \textit{et al.} \cite{durrieu2011musically} formulated an advanced signal model using multiple NMF decompositions. The predominant source is modeled with a source-filter model and all other sources are captured by an unconstrained NMF. First, the F0 of the predominant target source is estimated using the signal model. Then, the F0 is used to guide the separation. Nakamura and Kameoka \cite{nakamura2020harmonic} proposed a powerful signal model combining NMF and harmonic-temporal clustering and integrated a source-filter model. It allows for blind, learning-free separation of harmonic sounds. A drawback of NMF-based methods is the low degree of flexibility because only a fixed number of spectral templates is used to describe a signal. This limits their performance, especially when inherent assumptions are violated. 

Recently, efforts have been made to make more flexible deep learning based source separation also usable in cases where no mixture-target pairs are available for training. Most works focus on creating learning targets artificially from mixtures or side-information in order to train DNNs in a supervised way in the absence of real targets. Seetharaman \textit{et al.} \cite{seetharaman2020bootstrapping} obtain targets for singing voice/accompaniment separation by clustering time-frequency bins of mixtures using several simple perceptual cues. Hung \textit{et al.} \cite{hung2021transcription} obtain harmonic target masks from well-aligned musical scores and further support the training process using score transcription models. Also deep clustering models \cite{hershey2016deep} have been trained for speaker separation without ground truth signals \cite{drude2019unsupervised, tzinis2019unsupervised}. The targets are obtained by clustering the mixture based on spatial information. The methods above yield good results but require substantial amounts of (unlabeled) training data and cannot separate homogeneous correlated sources.






As an alternative, it has been proposed to train deep generative models on isolated source signals to use them subsequently for source separation \cite{narayanaswamy2020unsupervised} or speech enhancement \cite{leglaive2019semi}. However, this strategy is challenging for MSS because it requires a large amount of isolated source signals and uncorrelated sources.

Lastly, mixture invariant training has been proposed recently in \cite{wisdom2020unsupervised} and refined in \cite{wisdom2021sparse} for unsupervised learning of audio source separation without a need for artificial targets. During training, the sum of two mixtures is given as an input and the DNN has to separate all sources so that, given the respective optimal binary mixing matrices, the two mixtures can be reconstructed individually. Since it is necessary that the sources are uncorrelated \cite{wisdom2021sparse}, this approach is not an option for MSS.

The method proposed in this paper uses F0 information to separate the (possibly homogeneous) sources like the learning free-methods of \cite{ewert2012using, durrieu2011musically} and the supervised methods of \cite{petermann2020deep, gover2020score}. It provides better performance than learning-free methods and does not require expensive labeled data like supervised methods. Our learning strategy is fundamentally different from other unsupervised methods: it is not limited to uncorrelated sources like \cite{wisdom2021sparse} and does not rely on artificial source targets which require the availability of aligned scores \cite{hung2021transcription}, sufficient spatial information in the mixture \cite{tzinis2019unsupervised, drude2019unsupervised}, or non-homogeneous sources \cite{seetharaman2020bootstrapping}. The proposed training objective is to re-synthesize the mixture with differentiable parametric source models. The only assumptions are that the number of sources is known and that their F0s can be estimated. In contrast to the unsupervised methods reviewed above, the proposed one can separate homogeneous sources, requires only a small amount of unlabelled data, and provides interpretable and modifiable source estimates.

There is a recent line of research that explores the combination of data-driven and knowledge-based methods to take advantage of both paradigms \cite{monga2021algorithm, shlezinger2020model, engel2019ddsp}. The integration of differentiable source models in the DNN-based source separation process is inspired by this model-based deep learning research. Specifically related to our work are recent speech synthesis methods which use differentiable parametric voice models and estimate their parameters using DNNs \cite{wang2019neural, rao2020sfnet}. We use similar voice models but in a different context. Engel \textit{et al.} \cite{engel2019ddsp} implemented a code library for differentiable digital signal processing and show the advantages of model-based deep learning for tasks such as synthesis, timbre transfer and dereverberation. The DNN architectures and the differentiable signal processing implementations we use in our experiments are inspired by their work. To the best of our knowledge, the proposed method is the first one that uses model-based deep learning for MSS.



\section{Method}
\label{sec:method}

\begin{figure}[!t]
\centering
\includegraphics[width=3.4in]{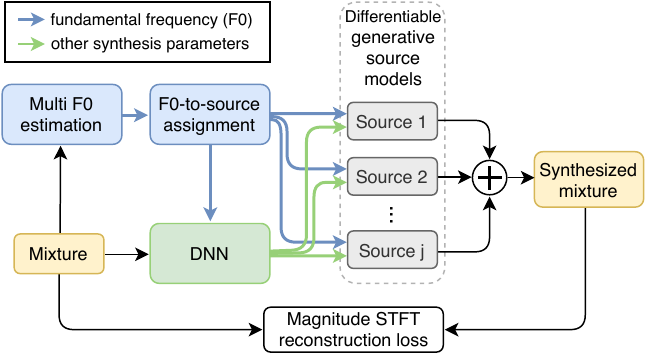}
\caption{Overview of the proposed unsupervised training procedure of a Deep Neural Network (DNN) for audio source separation.}
\label{fig:overview}
\end{figure}

We observe the single-channel mixture $m(t) = \sum_{j=1}^{J} s_j(t)$ of $J$ monophonic source signals $s_j(t)$ where $t~\in~\{1,...,T\}$ indexes discrete time samples. Our goal is to estimate all source signals $s_j$. We propose a novel approach to train a DNN for this task without access to any isolated source signals. The sources are modeled with differentiable parametric source models which we describe in Section~\ref{subsec:source_models}. The DNN estimates the source parameters given the F0 as explained in Section~\ref{subsec:param_estimation}. The objective of the unsupervised training strategy is to re-synthesize the mixture. Details are given in Section~\ref{subsec:training} and an overview of the procedure is presented in Fig. \ref{fig:overview}. At test time, the synthesized source signals can either be used directly as source estimates or soft masks can be derived from them for Wiener filtering of the mixture. Implementation details are described in Section~\ref{subsec:implementation}

\subsection{Source model}
\label{subsec:source_models}

\begin{figure*}[!t]
\centering
\includegraphics[width=7in]{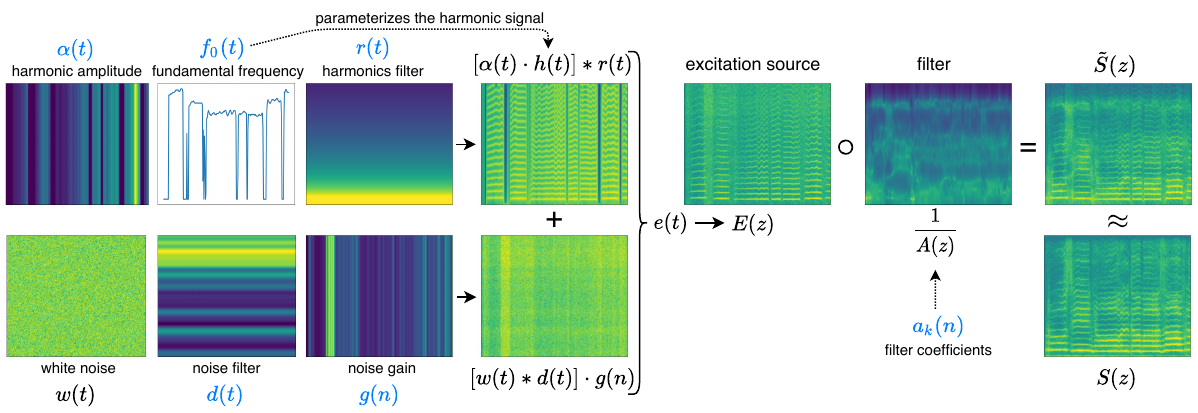}
\caption{Exemplary overview of the source-filter model decomposition. The model parameters are denoted in blue font. The '$\circ$' denotes element-wise multiplication. Although most components are visualized through magnitude spectrograms, processing is not necessarily done in the time-frequency domain.}
\label{fig:source_model}
\end{figure*}

The proposed method is not specific to any particular source model and any parametric model may be used as long as it can be formulated in a differentiable way. This is often facilitated by automatic differentiation software such as TensorFlow \cite{abadi2016tensorflow} or PyTorch \cite{paszke2019pytorch}. In this work, we use the source-filter model of speech production \cite{fant1970acoustic}. It describes a signal as an excitation signal from a sound source (e.g. the glottis) which is modified by a time-varying filter (e.g. the vocal tract) \cite{fant1970acoustic}. It is used to model a wide range of signals such as human voice \cite{fant1970acoustic, rao2020sfnet, degottex2013mixed} and musical instruments \cite{durrieu2011musically, heittola2009musical}. An exemplary visualization of our source-filter model is presented in Fig. \ref{fig:source_model}.

In the following, we assume that the true source signal $s_j(t)$ is segmented into $N$ frames of length $T'$ samples. The $n$-th frame is given by
\begin{equation}
    s_j(n,t) = s_j(t + nB), \hspace{5px} t \in \{1,...,T'\}
\end{equation}
where $B$ is the hop size between frames in samples and $n~\in~\{1,...,N \}$. We denote the estimate of the source signal frame generated by the source model using a tilde: $\tilde{s}_j(n,t)$. The source model may be formulated in the z-domain as
\begin{equation}
    \tilde{S}_j(n,z) = E_j(n,z) \frac{1}{A_j(n,z)}.
    \label{eq:filter_z_domain}
\end{equation}
$E_j(n,z)$ is the z-transform of the excitation signal $e_j(n,t)$ and $\frac{1}{A_j(n,z)}$ is the transfer function of a time-varying all-pole filter of order $K$. We drop the source index $j$ for brevity hereafter but we would like to emphasize that each source is modeled with its dedicated model. The filtering process in (\ref{eq:filter_z_domain}) is best described by the difference equation
\begin{equation}
    \tilde{s}(n,t) = e(n,t) - \sum_{k=1}^K a_k(n) \cdot \tilde{s}(n, t-k)
    \label{eq:difference_eq}
\end{equation}
where $a_k(n)$ are the filter coefficients for frame $n$ and~'$\cdot$' denotes scalar multiplication. We explain how to deal with frame boundaries and other implementation details in Section~\ref{subsec:implementation}.

A sinusoids plus noise model is employed to generate the excitation signal $e(n,t)$. It is an expressive synthesis model for music \cite{serra1990spectral} and speech signals \cite{laroche1993hnm, richard1996analysis, erro2013harmonics} which synthesizes sound as a sum of sinusoids and filtered white noise. A differentiable version was recently implemented by Engel \textit{et al.} \cite{engel2019ddsp, engel2020self} who showed impressive results using it for model-based deep learning. Since we model a monophonic source, we constrain the sinusoid frequencies to be integer multiples of a fundamental frequency. The model thus reduces to the \emph{harmonics plus noise} model \cite{laroche1993hnm, engel2019ddsp} which we formulate as
\begin{equation}
    e(n,t) = [\alpha(n,t) \cdot h(n,t)] \ast r(t) + [w(t) \ast d(t)] \cdot g(n)
\end{equation}
where $\ast$ denotes the convolution operator, $\alpha(n,t)$ is the time-varying amplitude of the harmonic signal $h(n,t)$, and $r(t)$ and $d(t)$ are Impulse Responses (IR) of time-invariant finite impulse response (FIR) filters. $w(t)$ is a uniform white noise signal and $g(n)$ is the constant noise gain for frame $n$.

The harmonic signal $h(n,t)$ is defined as
\begin{equation}
        h(n,t)~=~\sum_{i=1}^I \sin(\phi_i(n,t))
    \end{equation}
    \begin{equation}
        \phi_i(n,t) = 2 \pi \sum_{v=1}^t i \cdot f_0(n,v) / f_s
        \label{eq:inst_phase}
    \end{equation}
where $\phi_i$ is the instantaneous phase of the $i$-th harmonic, $f_0$ is the fundamental instantaneous frequency, and $f_s$ is the sampling frequency. The initial phase is assumed to be zero. Equation (\ref{eq:inst_phase}) is a numerical approximation of integration based on \textit{sample and hold} \cite[Ch.~4]{horowitz2002art}. Note that the signal $h(n,t)$ is fully parameterized by the time-varying fundamental frequency $f_0$.

The filter $r(t)$ imposes a fixed spectral shape on $h(n,t)$. Without $r(t)$, all sinusoids have the same amplitude. However, for certain sound sources a specific time-invariant spectral shape can be assumed, e.g. the spectral roll-off of the glottal signal \cite{fant1970acoustic}. Alternatively, a specific amplitude parameter may be used for each sinusoid in $h(n,t)$ \cite{serra1990spectral, laroche1993hnm}. However, we choose to make the gain dependent on the frequency and not on the harmonic number. Similarly, $d(t)$ determines the spectral shape of the noise component. Both filters are time-invariant so that they only account for the global spectral shape. Short term variations, e.g. due to articulations of words, are modeled by the all-pole filter $\frac{1}{A(n,z)}$.

The source model parameters are $\{ a_k(n), \alpha(t), f_0(t), r(t), g(n), d(t)\}$. In the next section, it is explained how they are obtained. $\alpha$ and $f_0$ need to vary slowly enough over time for the model to be mathematically identifiable. This is indirectly enforced by the way these parameters are estimated which leads to smooth trajectories.


\subsection{Parameter estimation}
\label{subsec:param_estimation}



We assume that the fundamental frequencies for each of the $J$ sources can be obtained from the mixture signal with a multiple F0 estimation system. Given that many such systems exist \cite{klapuri2006multiple, cuesta2020multiple, zhang2020multi} and that it is still an active research area, we are confident that this is a reasonable assumption. When all F0s are obtained, each F0 value needs to be assigned to one specific source. Various solutions for the F0-to-source assignment problem have been proposed \cite{chew2004separating, mcleod2016hmm, schramm2017multi}. Most of them are based on principles such as temporal pitch continuity, low voice crossing probability, and minimal temporal gaps within a voice \cite{mcleod2016hmm}. In our experiments we use a heuristic based on these principles, cf. Section \ref{subsec:experiment_setup}. F0 estimates are usually provided at a frame rate which is smaller than the sample rate \cite{klapuri2006multiple, cuesta2020multiple, zhang2020multi}. Therefore, following \cite{engel2019ddsp}, the source specific F0 time series are upsampled to the sample rate using bilinear interpolation. This leads to smooth trajectories.

In the following, we describe how the remaining synthesis parameters are estimated with a DNN for each source given its F0. The task the DNN has to solve is similar to the one of NMF in the context of learning-free F0-informed source separation in \cite{ewert2012using, durrieu2011musically}. Note that the differentiable source models do not put any constraints on the neural network type or architecture which is used to estimate the parameters. Here we use a simple DNN as in \cite{engel2019ddsp} and focus on the advantages of including parametric source models in deep learning based separation. 

\begin{figure}[!t]
\centering
\includegraphics[width=3.2in]{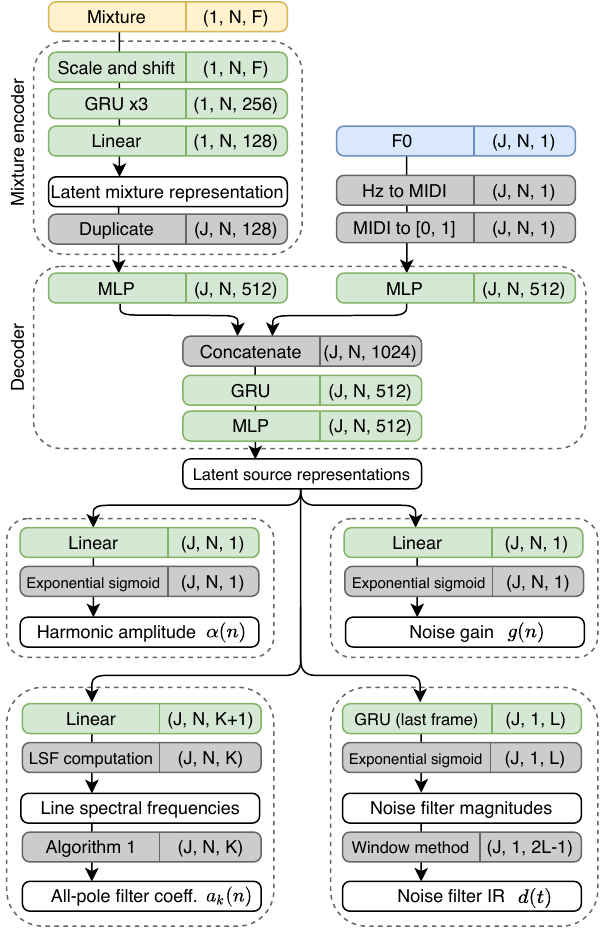}
\caption{Overview of the processing steps for the parameter estimation. Transformations with learnable parameters are shown in green, predefined processing steps in gray, (intermediate) outputs in white boxes. The output shape of a transformation is shown in the right part of the box.}
\label{fig:param_estimation}
\end{figure}

The mixture signal is represented by the logarithmic magnitude of its spectrogram obtained by a Short Time Fourier Transform (STFT) of $m(t)$. The spectrogram has $F$ frequency bins and $N$ time frames. Each spectrogram is normalized by subtraction of its mean and division by its standard deviation. Then, each frequency bin is scaled and shifted by dedicated learned scalars. The DNN architecture is similar to the one used in \cite{engel2019ddsp}. An overview of the DNN and further processing steps for the parameter estimation is presented in Fig. \ref{fig:param_estimation}. We use linear layers and unidirectional Recurrent Neural Networks (RNN) with Gate Recurrent Units (GRUs) \cite{cho2014properties}. The Multi-Layer Perceptron (MLP) consists of three repetitions of linear layer, layer normalization \cite{ba2016layer}, Leaky ReLU activation \cite{maas2013rectifier}.

The mixture encoder learns a latent representation of the mixture and then creates as many duplicates as there are sources. Each latent mixture copy is then combined with the F0 information of one source by the decoder. The F0 is provided at the frame rate of the mixture STFT. The F0 values are converted from Hertz to MIDI note numbers which are then normalized to the interval $[0, 1]$. The decoder computes a separate latent representation for each source. The source model parameters are obtained from this source representation by one last transformation with learned parameters (linear layer or GRUs) followed by some predefined processing steps. 
The frame-wise harmonic amplitude $\alpha(n)$ and the noise gain $g(n)$ are computed with a linear layer with an exponential sigmoid activation function \cite{engel2019ddsp} defined as
\begin{equation}
    y = y_{\text{max}} \cdot \text{sigmoid}(x)^{\log(10)} + 10^{-7}
    \label{eq:exp_sigmoid}
\end{equation}
where $x$ and $y$ are the input and output value, respectively, and $y_{\text{max}}$ is a scalar determining the upper bound of $y$.
Following \cite{engel2019ddsp}, the harmonic amplitude is then upsampled to the sample rate using overlapping Hann windows which yields a smooth $\alpha(t)$. The noise gain is only required at frame rate.

The filter with impulse response $d(t)$ is time-invariant. Therefore, the network output from which $d(t)$ is computed should summarize information about the whole source signal. We obtain such an output by processing the latent source representation with a unidirectional RNN with GRUs and then using only the output at the last time frame for further processing. This last output frame is processed with the exponential sigmoid presented in (\ref{eq:exp_sigmoid}) which results in a tensor of shape $(J, 1, L)$. The tensor contains $L$ samples of the magnitudes of the single-sided frequency responses of the noise filters for $J$ sources. The samples define a zero-phase FIR filter according to the frequency sampling method \cite{smith2011frequency}. Using the window method \cite{smith2011generalized}, we obtain the impulse response $d(t)$ as it is also done in \cite{engel2019ddsp}.

The impulse response $r(t)$ of the time-invariant harmonics filter can be obtained in the same way as $d(t)$ from a DNN output. One may also wish to make the filters time-varying by using a linear layer for the last transformation or using all GRU outputs. However, for the scope of this work, we fix $r(t)$ manually. More details about $r(t)$ are given in Section~\ref{subsec:experiment_setup} where we describe the experimental setup. 

For the estimation of the parameters we addressed so far, practical ways have already been proposed by Engel \textit{et al.} \cite{engel2019ddsp}. More care needs to be taken when obtaining Infinite Impulse Response (IIR) filters such as $\frac{1}{A(z)}$ from DNN outputs because it must be avoided that the filter becomes unstable. The filter $\frac{1}{A(z)}$ of order $K$ is fully defined by the filter coefficients $a_k$ with $k\in \{1,...,K\}$ (see also the difference equation in (\ref{eq:difference_eq})). However, no condition which guarantees stability can be formulated for the filter coefficients directly.

Different parameterizations of all-pole filters exist which allow for the formulation of stability criteria. One option would be to estimate $K$ reflection coefficients \cite{rabiner2010theory} with the DNN. Stability is guaranteed if the coefficients are within the interval $]-1, 1[$. They can be converted to the filter coefficients with a simplified version of the Levinson-Durbin algorithm \cite{levinson1946wiener, durbin1960fitting}, see also \cite{rabiner2010theory}. This approach was used in \cite{rao2020sfnet} to define the all-pole vocal tract filter with a DNN for speech synthesis. The drawback of this method is that conclusions about the filter's frequency response can neither be drawn from the reflection coefficients nor the filter coefficients.

Therefore, we choose to parameterize the all-pole filter with Line Spectral Frequencies (LSFs) \cite{itakura1975line}. LSFs are related to the positions of the filter poles and thus to the frequency response \cite{rabiner2010theory}. Hence, they provide an interpretable parametrization. They also allow the formulation of constraints to control the filter response and can be interpolated \cite{chu2004speech}. LSFs were introduced in \cite{itakura1975line} as an alternative representation of linear prediction coefficients. Below, we briefly explain their definition and how we use them. For a comprehensive overview of LSFs, we refer the reader to \cite{chu2004speech, soong1984line, kabal1986computation}.

The polynomial $A(z) = 1 - \sum_{k=1}^K a_k z^{-k}$ can be decomposed into the symmetric and antisymmetric polynomials $P(z)$ and $Q(z)$ of order $K+1$ as
\begin{equation}
    A(z) = \frac{P(z) + Q(z)}{2}.
\end{equation}
It can be shown that if the roots of $P(z)$ and $Q(z)$ alternate on the unit circle, the corresponding filter $\frac{1}{A(z)}$ is stable and minimum-phase \cite{soong1984line}. The unit circle in the z-plane is described by $z=e^{-j\omega}$ where $\omega$ is the phase angle in radiants. Hence, $\omega$ describes the location of the roots. If $K$ is even, $P(z)$ has a root at $z=-1$ and $Q(z)$ has a root at $z=+1$. The remaining roots occur in complex conjugate pairs. Therefore, it is sufficient to consider only the roots on the upper semicircle. The angles $\omega_k$ defining the locations of these complex roots are called LSFs. Two to three LSFs tend to be close together when a filter pole is close to the unit circle in their proximity which corresponds to a peak in the frequency response, hence their frequency domain interpretation. If $K$ is even, $P(z)$ and $Q(z)$ have $K/2$ complex roots on the upper unit semicircle each, for which the following relation holds:
\begin{equation}
    0 < \omega_k < \omega_{k+1} < \pi.
\label{eq:lsf_inequality}
\end{equation}
When $k$ is odd, $\omega_k$ defines a root of $P(z)$; when $k$ is even, it defines a root of $Q(z)$ for $k \in \{1,...,K \}$. In other words, a stable minimum-phase filter $\frac{1}{A(z)}$ of order $K$ is defined by $K$ LSFs fulfilling the relation in (\ref{eq:lsf_inequality}).

We obtain such LSFs as follows. The latent source representations are transformed by a linear layer which yields a tensor of shape $(J,~N,~K+1)$. It is processed by an exponential sigmoid activation with $y_{max}~=~2$. The resulting tensor can be viewed as $J \cdot N$ vectors $\mathbf{v}~\in~\mathbb{R}^{K+1}$. The vectors are normalized so that their entries $v_k$ sum up to $\pi$:
\begin{equation}
    \mathbf{\bar{v}} = \frac{\mathbf{v}}{\sum_{k=1}^{K+1} v_k} \cdot \pi.
\end{equation}
The $K$ LSFs respecting (\ref{eq:lsf_inequality}) are then obtained by the cumulative sum
\begin{equation}
    \omega_k = \sum_{i=1}^k \bar{v}_i \hspace{5px} \text{for} \hspace{5px} k = 1,...,K.
\end{equation}
Finally, the LSFs can be converted to filter coefficients using Algorithm \ref{alg:lsf_to_lpc} \cite{mcloughlin2008line},\footnote{The formulation of Algorithm \ref{alg:lsf_to_lpc} which we present in this paper has been presented in \cite{mcloughlin2008line}. Some equations in the main body of \cite{mcloughlin2008line} contain errors but the Matlab code in the Appendix is correct. A less general formulation is found in \cite[Ch.~8]{chu2004speech}. The conversion was formally introduced in \cite{kabal1986computation}.} \cite{chu2004speech, kabal1986computation}.

\begin{algorithm}
    \caption{Compute filter coefficients $a_k$ from $\omega_k$ \cite{kabal1986computation, mcloughlin2008line}}
    \begin{algorithmic}
    \REQUIRE $(\omega_k)_{k=1:K}$
    \STATE \textbf{Define:} $x_k = \cos(\omega_k)$
    \STATE \textbf{Initialize:} $p_{-1}' = q_{-1}' = 0;$ \hspace{5px} $p_0' = q_0' = 1$
    \STATE \textbf{Initialize:} $p_1' = - 2 x_{1};$ \hspace{5px} $q_1' = -2x_2$
    \FOR{$k=2$ \TO $K/2$}
        \STATE $p_k' = - 2 p_{k-1}' x_{2k-1} + 2 p_{k-2}'$
        \STATE $q_k' = - 2 q_{k-1}' x_{2k} + 2 q_{k-2}'$
            \FOR{$i=(k-1)$ \TO $1$}
            \STATE $p_i' = p_i' - 2 p_{i-1}' x_{2k-1} + p_{i-2}'$
            \STATE $q_i' = q_i' - 2 q_{i-1}' x_{2k} + q_{i-2}'$
            \ENDFOR
    \ENDFOR
    \FOR{$k=1$ \TO $K/2$}
    \STATE $p_k = p_k' + p_{k-1}'$
    \STATE $q_k = q_k' - q_{k-1}'$
    \ENDFOR
    \FOR{$k=1$ \TO $K/2$}
    \STATE $a_k = (p_k + q_k)/2$
    \STATE $a_{(K/2 + k)} = (p_{(K/2 - k + 1)} - q_{(K/2 - k + 1)})/2$
    \ENDFOR
    \ENSURE $(a_k)_{k=1:K}$
    \end{algorithmic}
    \label{alg:lsf_to_lpc}
\end{algorithm}

To sum up the parameter estimation, F0s are estimated from the mixture and assigned to the sources using existing methods. $a_k(n)$, $\alpha(t)$, $g(n)$, and $d(t)$ are obtained with a DNN and $r(t)$ is fixed manually in this work but may also be estimated by a DNN.

\subsection{Unsupervised training}
\label{subsec:training}

The proposed training procedure requires only mixture signals, no isolated source signals are needed. During training, the task of the DNN is to reconstruct the observed mixture by estimating the corresponding parameters of the source models. A schematic overview of the training process is presented in Fig.~\ref{fig:overview}. The generated mixture estimate $\tilde{m}(t)$ is the sum of the source signals generated by the source models:
\begin{equation}
    \tilde{m}(t) = \sum_{j=1}^J \tilde{s}_j(t).
\end{equation}
In theory, the source models make it possible to synthesize a mixture estimate $\tilde{m}(t)$ which is perceptually identical to the true mixture $m(t)$. Since absolute phase offsets are irrelevant for human perception, the true and estimated mixtures do not need to have the same phase. Therefore, the reconstruction loss $\mathcal{L}_{rec}$ is formulated as a multi-scale spectral loss \cite{engel2019ddsp}
\begin{equation}
     \mathcal{L}_c = \Vert \mathbf{M}_c - \mathbf{\tilde{M}}_c \Vert_1 + \Vert \log(\mathbf{M}_c) - \log(\mathbf{\tilde{M}}_c) \Vert_1
 \end{equation}
\begin{equation}
    \mathcal{L}_{rec} = \sum_c \mathcal{L}_c
\end{equation}
where $\mathbf{M}_c$ and $\mathbf{\tilde{M}}_c$ denote the magnitude spectrograms of the input mixture and its estimate, respectively, and $c~=~[2048, 1024, 512, 256, 128, 64]$ indicates the FFT size used to compute the STFT. The frames overlap by 75\%.

The separation of the sources is essentially ensured by the assignment of the F0s to the sources similar to score/F0-informed separation with NMF \cite{ewert2012using, durrieu2011musically}. The DNN has to estimate the remaining parameters for each source in order to minimize the loss.

At test time, the DNN parameters are fixed and a soft mask for source $j$ is obtained by the element-wise division $\mathbf{\tilde{S}}_j / \sum_{j=1}^{J} \mathbf{\tilde{S}}_j$ where $\mathbf{\tilde{S}}_j$ is the magnitude spectrogram of the generated source signal~$\tilde{s}_j$. The \textit{final} time-domain source estimates, marked with a hat, $\hat{s}_j$ are obtained by Wiener filtering using the soft masks.

\subsection{Implementation details}
\label{subsec:implementation}

We implemented the proposed method using the PyTorch framework \cite{paszke2019pytorch}. For the differentiable source models, we make use of the DDSP library \cite{engel2019ddsp}. We re-implemented it in PyTorch and added extensions such as Algorithm \ref{alg:lsf_to_lpc} and an all-pole filter. The code is available online\footnote{\url{https://github.com/schufo/umss}}.

Using an all-pole filter in the proposed framework entails two challenges. Firstly, the autoregressive filtering process is slow because it does not allow for precise parallel processing of frames. Secondly, the filter is time-varying, i.e. its coefficients are different at every frame. Therefore, extra care must be taken to ensure a smooth transition between frames to avoid artefacts. The DNN operates at a frame rate which is determined by the FFT size $T'$ and hop size $B$ used to compute the STFT of the mixture. Hence, the DNN estimates a set of $K$ filter coefficients for each frame. We apply the all-pole filter to all frames in parallel using the difference equation in (\ref{eq:difference_eq}) in order to make filtering faster. The initial states $\tilde{s}(n, t)$ with $t~\le~0$ are set to zero for each frame. The output frames are then multiplied with a Hann window and the final output signal is obtained by the overlap-add method. It is therefore important that the hop size $B$ is chosen so that the Hann window respects the constant overlap-add condition. We use $B~=~T'~/~2$ in our experiments. Windowing and 50\% overlap make the transition between frames smooth. The errors that are introduced by setting the initial states to zero instead of taking samples of the previous frame into account (which is not possible in parallel processing) are negligible: Firstly, the errors are larger at the start of each frame where their importance is mitigated by the window. Secondly, since the filter coefficients are different at each frame, the importance of samples from the previous frame is reduced.

We found it to be critical to implement Algorithm \ref{alg:lsf_to_lpc} with double precision (64-bit floating point) because it is more sensitive to rounding errors with increasing filter order, which can lead to unstable filters.

The excitation signal $e(t)$ is computed as follows. The harmonic component $\alpha(t) 
\cdot h(t)$ and the noise $w(t)$ are generated in the time domain for the entire signal length $T$. The time-invariant FIR filters $r(t)$ and $d(t)$ and the noise gain $g(n)$ are applied frame-wise in the frequency domain followed by overlap-add.

\section{Experiments}
\label{sec:experiments}

We evaluate the proposed approach on an \textit{a cappella} vocal ensemble separation task. The goal is to estimate the individual signals of $J$ singers from their mixture. This task is a good choice for evaluation because sources in vocal ensembles are homogeneous and correlated. Moreover, singing voice is a challenging musical source. It has a strongly time-varying spectral envelop and also produces sounds without any harmonic content such as unvoiced consonants. Also, only small amounts of data for supervised training are available for vocal ensemble separation. This makes unsupervised learning an important alternative.


\subsection{Data}
As training and validation data, we use the Bach Chorals (BC) dataset\footnote{\url{https://www.pgmusic.com/bachchorales.htm}} and the Barbershop Quartet (BQ) dataset\footnote{\url{https://www.pgmusic.com/barbershopquartet.htm}}. The BC set contains 26 chorals sung by a vocal quartet with the voices Soprano, Alto, Tenor, Bass (SATB). The BQ set contains 22 songs performed by a vocal quartet comprising the voices tenor, lead, baritone and bass. All voices are available in isolation for both sets. This allows us to compare the proposed unsupervised approach to supervised baselines.

We combine the BC and BQ sets to generate what we call the \textit{full} training and validation sets. The \textit{full} validation set comprises songs 8 and 9 of the BC set and songs 8 and 9 of the BQ set and has a total length of 9 minutes and 10 seconds. The remaining songs build the \textit{full} training set with a total length of 91 minutes and 20 seconds. We also build a \textit{small} training set consisting of BC song 1 with a length of 2 minutes and 40 seconds and a \textit{small} validation set consisting of BC song 2 with a length of 2 minutes and 20 seconds. When mixtures with less than four singers are created from the individual voice recordings, all possible combinations of the four voices with the desired number of singers are used with the constraint of using only one singer per voice.

As test data, we use the Choral Singing Dataset \cite{cuesta2018analysis}. It comprises three songs performed by an SATB choir with four singers per voice. All 16 singer signals are available in isolation which allows to evaluate the separation with objective metrics. We add the signals of individual singers (max. one per voice) to produce the test mixtures. For mixtures of $J=4$ singers, the test set has a length of 6 minutes and 48 seconds. For mixtures of $J=2$ singers, the test set has a length of 40 minutes and 48 seconds due to more possible voices combinations.

We resample the training, validation, and test data to a sample rate of 16 kHz. The training examples are excerpts of 4 seconds length which are randomly drawn from the training set. The validation and test set are split into fixed excerpts of 4 seconds length. There is no overlap regarding singers, songs, or recording setup between the test and training data. While the training data contain a considerable amount of reverberation, the test recordings are much less reverberant.

\subsection{Experimental setup}
\label{subsec:experiment_setup}

We perform two sets of experiments: one using mixtures of $J=2$ singers for training and testing, and the other using mixtures of $J=4$ singers.

The F0s are obtained from the mixture signals using the multiple F0 estimation model of Cuesta \textit{et al.} \cite{cuesta2020multiple}. We use the pre-trained "Model 3" which is available online\footnote{\url{https://github.com/helenacuesta/multif0-estimation-polyvocals}}. For the F0-to-source assignment on the given data, we found that a simple heuristic is sufficient. It is based on the same principles as more advanced solutions such as temporal  pitch  continuity,  low  voice crossing probability, and minimal temporal gaps within a voice  \cite{mcleod2016hmm, schramm2017multi}. The F0 estimator provides $x$ F0 values at each time frame. First, we process all frames where $x~=~J$. The F0 values are sorted according to magnitude and assigned to the voices assuming they do not cross. Subsequently, the remaining frames are processed. When $x~<~J$ we assume that some voices are silent. We assign each F0 value to the source which has the closest F0 value in a previous or subsequent frame (pitch continuity principle). The zero value is assigned to silent sources. In the rare case that $x>J$, we sort the values according to magnitude and select $J$ F0s using the pitch continuity principle and assign them to the sources.

The mixture spectrograms are computed using an FFT size of $T'=512$ and a hop size of $B=256$ samples. Hence, they have $F=257$ frequency bins and $N=250$ time frames. We fix the impulse response $r(t)$ so that the frequency response of the FIR filter falls off with a rate of 6 dB/octave, with a reference frequency of 200 Hz below which the response is flat. We chose this rate because it accounts for the combined spectral characteristics of the glottal source and lip radiation \cite{fant1970acoustic}. Estimating $r(t)$ with the DNN instead did not lead to improvements. We set the order of the all-pole filter to $K=20$. The spectrograms of the synthesized source signals $\mathbf{\tilde{S}}_j$ to compute the soft masks are computed with an FFT size of 2048 and a hop size of 256 samples.

Training is done with the ADAM optimizer \cite{kingma2014adam}, a batch size of 16 and a learning rate of 0.0001. Training is stopped after 200 consecutive epochs without improvement of the validation loss.

We train the model with the proposed unsupervised approach on the \textit{full} and on the \textit{small} training set. We call the experiments UnSupervised-Full (US-F) and UnSupervised-Small (US-S), respectively. As a reference, we also train the same model in a supervised way on the same data. In this case, the loss is computed for each source estimate individually using its target. The total loss is the sum of the "source losses". We call these experiments SuperVised-Full (SV-F) and SuperVised-Small (SV-S).

Since the proposed method is dependent on available F0 estimations, we also evaluate its robustness to F0 estimation errors. The corresponding experiments and results are explained in Section \ref{subsec:f0_robustness}.

\subsection{Baselines}

We compare the proposed unsupervised approach to two learning-free methods and one supervised approach. The baselines also exploit F0 information and compute soft masks for Wiener filtering.

The first learning-free baseline was proposed by Ewert and Müller \cite{ewert2012using}. It approximates the mixture magnitude spectrogram with a simple NMF decomposition:
\begin{equation}
    \mathbf{M} \approx \mathbf{\hat{M}} = \mathbf{W} \mathbf{H}
\end{equation}
where $\mathbf{W} \in \mathbb{R}^{F\times R}$ is a matrix of $R$ spectral templates and $\mathbf{H}~\in~\mathbb{R}^{R\times N}$ contains their activations over $N$ time frames. In \cite{ewert2012using}, $\mathbf{W}$ and $\mathbf{H}$ are initialized using information from an aligned musical score. One spectral template per semitone is used. In our experiments, we have F0 information available, which is more precise than a semitone scale. Therefore, we use a scale with a precision of $\frac{1}{10}$ of a semitone. The F0 values are converted from Hertz to MIDI numbers which are rounded to one decimal place for this purpose. The F0s are used for initialization and for the separation to determine which activations belong to which source. After testing different combinations, we obtained the best results with an FFT size of 2048 and a hop size of 256 samples to compute the spectrograms. We call this method NMF1.

The second learning-free baseline is the method proposed by Durrieu \textit{et al.} \cite{durrieu2011musically}. The target source is modeled with a source-filter model and the residual sources are modeled with a conventional NMF. The method approximates the power spectrogram of the mixture $\mathbf{M}_{\text{pow}}$ as
\begin{equation}
    \mathbf{M}_{\text{pow}} \approx \mathbf{\hat{M}}_{\text{pow}} = \underbrace{(\mathbf{W}^{\Gamma} \mathbf{H}^{\Gamma} \mathbf{H}^{\Phi})}_{\text{filter}} \circ \underbrace{(\mathbf{W}^{F0} \mathbf{H}^{F0})}_{\text{source}} + \underbrace{(\mathbf{W}^{O} \mathbf{H}^{O})}_{\text{residual}}
    \label{eq:durrieu}
\end{equation}
where $\circ$ denotes element-wise multiplication. $\mathbf{W}^{\Gamma} \in \mathbb{R}^{F \times P}$ contains $P$ spectral atoms consisting of shifted Hann windows with 75\% overlap so that the whole frequency range is covered across $\mathbf{W}^{\Gamma}$. The matrix $\mathbf{H}^{\Gamma} \in \mathbb{R}^{P \times K}$ contains their activations to combine them to smooth filters and $\mathbf{H}^{\Phi} \in \mathbb{R}^{K \times N}$ contains activations to combine the smooth filters. $\mathbf{W}^{F0} \in \mathbb{R}^{F \times U}$ contains a fixed set of $U$ spectral templates defined by the glottal source model KLGLOTT88 \cite{klatt1990analysis}. There is one spectral template for each F0 in steps of $\frac{1}{20}$ semitone between a minimum and a maximum frequency. $\mathbf{H}^{F0} \in \mathbb{R}^{U \times N}$ contains the activations of the spectral templates. In \cite{durrieu2011musically}, $\mathbf{H}^{F0}$ is initialized using F0 information of the predominant source estimated using the signal model in (\ref{eq:durrieu}). We initialize $\mathbf{H}^{F0}$ using the F0 information we obtained from the multi-pitch estimation \cite{cuesta2020multiple}. In \cite{durrieu2011musically}, the spectral templates of the residual sources $\mathbf{W}^O \in \mathbb{R}^{F \times R}$ and their activations $\mathbf{H}^O \in \mathbb{R}^{R \times N}$ are initialized randomly. We initialize them using the F0 information for the corresponding sources as done in NMF1. This leads to improvements. We call this baseline NMF2. The parameters to be estimated are $\{\mathbf{H}^{\Gamma}, \mathbf{H}^{\Phi}, \mathbf{H}^{F0}, \mathbf{W}^{O} \mathbf{H}^{O} \}$. For NMF2, we obtained the best results using an FFT size of 1024 and a hop size of 128 samples. To the best of our knowledge, these two baselines are among the best learning-free, informed methods for musical and homogeneous source separation. 

Furthermore, we train the F0-informed supervised deep learning approach for vocal ensemble separation proposed by Petermann \textit{et al.} \cite{petermann2020deep} on our data. They use a classical U-Net architecture with a control mechanism \cite{meseguer2019conditioned}. The F0 information is used to select the target source and to guide the separation. For this baseline, mixture and target source spectrograms are computed using an FFT size of 1024 and a hop size of 256 samples. Wiener filtering is applied at test time using all $J$ source estimates to compute soft masks.
It is trained with the ADAM optimizer \cite{kingma2014adam}, a batch size of 16 and a learning rate of 0.001. We train this baseline on the full and the small training set and call the experiments Unet-F and Unet-S, respectively. 
\color{black} Note that all baselines make use of the F0 information for the separation.
\color{black}

\section{Results and discussion}
\label{sec:results_discussion}

\begin{figure*}[!t]
\centering
\subfloat[$J=2$ sources]{\includegraphics[width=3.24in]{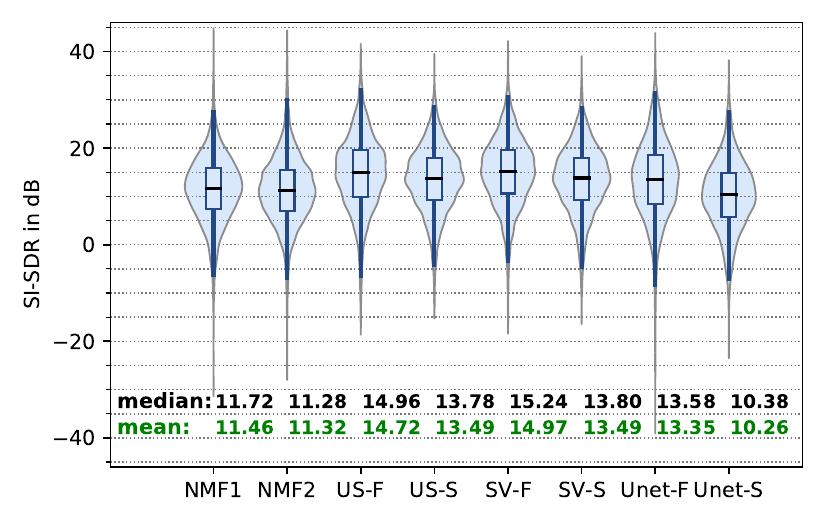}%
\label{fig:si_sdr_j=2}}
\hfil
\subfloat[$J=4$ sources]{\includegraphics[width=3.28in]{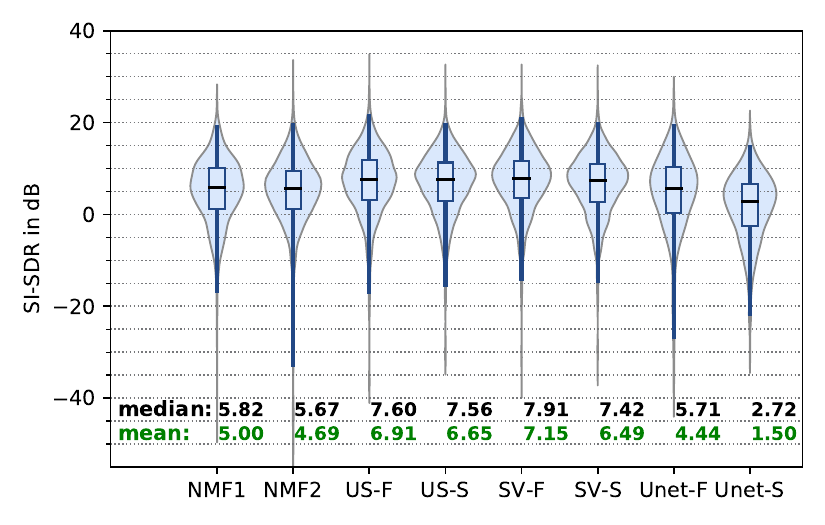}%
\label{fig:si_sdr_j=4}}
\caption{Violin plots and box plots of the SI-SDR values in dB for all evaluation frames. The boxes extend from the first to the third quartile, the medians are marked with a black horizontal line. The box plot whiskers (dark blue) extend from the first to the 99th percentile. The violin plots extend over the whole data range. In (b), NMF2 has five outliers between -60 and -80 dB which are not shown.}
\label{fig:si_sdr}
\end{figure*}

\begin{figure*}[!t]
\centering
\subfloat[$J=2$ sources]{\includegraphics[width=3.44in, trim=12 25 13 15, clip]{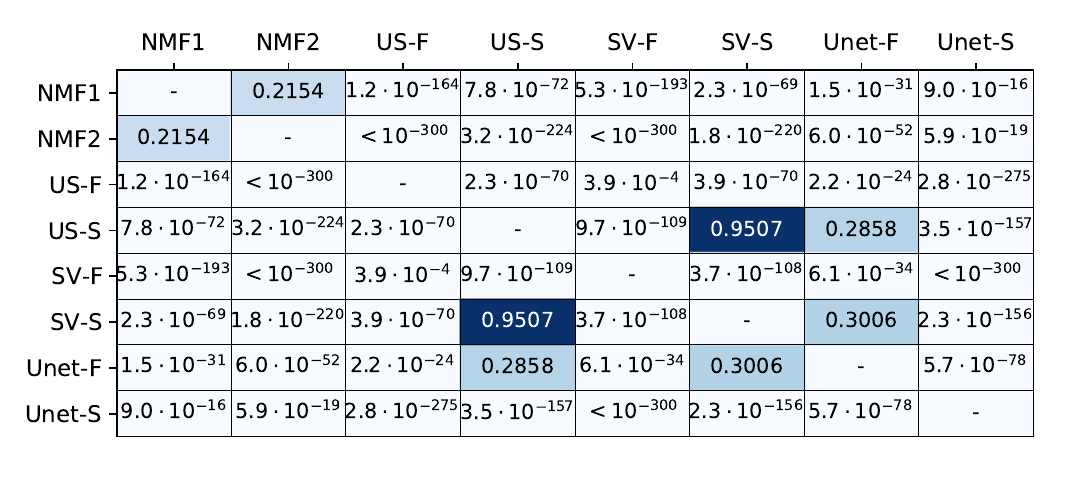}%
\label{fig:p_values_j=2}}
\hfil
\subfloat[$J=4$ sources]{\includegraphics[width=3.44in, trim=12 25 13 15, clip]{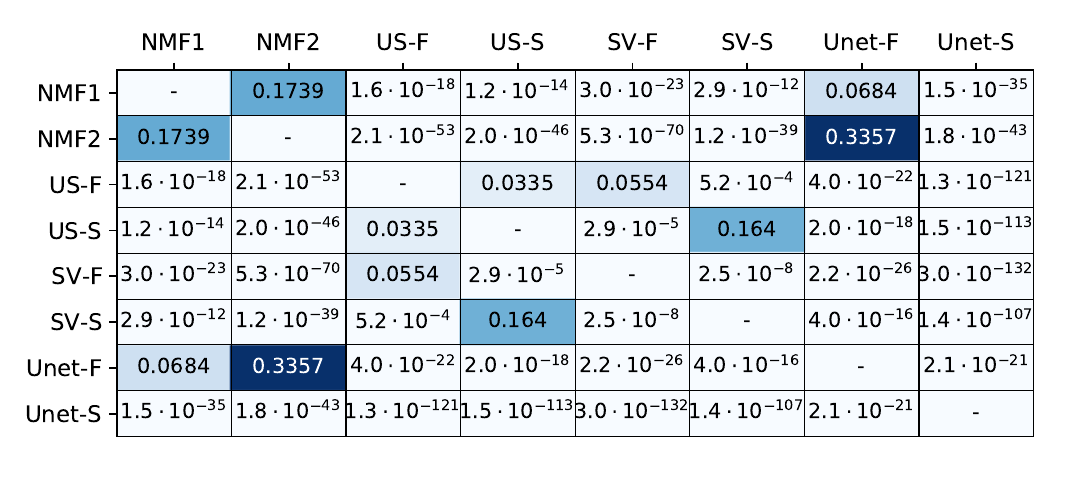}%
\label{fig:p_values_j=4}}
\caption{The p-values of pair-wise t-tests between the distributions of SI-SDR values for all experiments.}
\label{fig:p_values}
\end{figure*}

\subsection{Experimental results}
\label{subsec:exp_results}
The separation quality was evaluated using the objective metric Scale-Invariant Source-to-Distortion ratio (SI-SDR) \cite{le2019sdr}. It is computed on evaluation frames of one second length without overlap as usually done for musical source separation evaluation \cite{stoter20182018}. The results for the cases of $J=2$ and $J=4$ sources are shown in Fig. \ref{fig:si_sdr} (a) and (b), respectively. The data points for the box plots and violin plots are the SI-SDR values in dB for all evaluation frames in the test set.
Target source frames, in which the sum of squares of the samples is below a threshold of 10, are considered to be silent and thus excluded from the evaluation.
For methods in which random numbers are involved, the evaluation was run with five different seeds to initialize the pseudorandom number generator. These methods are NMF2 (random initialization of $\mathbf{H}^{\Gamma}$ and $\mathbf{H}^{\Phi}$) and the proposed approach (random white noise) used in experiments US-F, US-S, SV-F, and SV-S.

We conducted two-sided t-tests \cite{moore2007basic} to assess whether the means of the SI-SDR score distributions are significantly different for each pair of experiments in our study. We used a Levene test \cite{levene1961robust} to assess whether a pair of SI-SDR score distributions has the same variance or not. If true, the comparison was made with a Student's t-test. If false, Welch's t-test \cite{welch1947generalization} was used. The resulting p-values \cite{moore2007basic} are shown in Fig. \ref{fig:p_values} (a) and (b) for $J=2$ and $J=4$, respectively. Most p-values are extremely small being in the order of $10^{-4}$ or smaller. This indicates that the corresponding means are significantly different. It can be seen that a few p-values are considerably larger. In this case it is more likely that the true means are not different.

In general, the SI-SDR is higher for the separation of mixtures of two sources compared to the four sources case. However, the relative performance of the methods is the same for both cases with the exception that Unet-F outperforms NMF1 and NMF2 when $J=2$ but not when $J=4$. This is related to the small amount of training data for a supervised deep learning model. Listening examples are available online\footnote{\url{https://schufo.github.io/umss/}}.

The proposed unsupervised method (US-F, US-S) performs better than the baselines. 
Its performance is very close to the one which is reached by the same model trained in a supervised way: \mbox{SV-F} is only slightly better than \mbox{US-F}, while SV-S and US-S have the same performance (p-values of 0.9507 and 0.164 for $J=2$ and $J=4$). This means that the proposed method achieves almost the same performance whether isolated target sources are available for training or not. This can be explained by the fact that the F0 information is used very efficiently by the proposed method. The F0 fully parameterizes the harmonic source component $h(t)$ and, hence, defines the corresponding source to a large extent. The DNN has to determine the remaining parameters which, given the F0, can be inferred from the mixture. Hence, isolated source targets do not carry major additional information.

Another interesting observation is that the performance of the proposed method does not drop drastically when the amount of training data is decreased by 97\% (US-F vs. US-S and SV-F vs. SV-S). For $J=2$, a decrease in SI-SDR can be seen but it is smaller than for the supervised baseline (Unet-F vs. Unet-S). For $J=4$, the performance difference of the proposed approach is very small when comparing training on the full and the small training set. For the unsupervised version the difference is probably not significant since the p-value of 0.0335 for the comparison of US-F and US-S is larger than most other p-values. In contrast, the SI-SDR of the Unet baseline drops strongly for $J=4$ as well. This shows that it is beneficial to integrate domain knowledge in the form of explicit source models in the separation model. 

We believe that the main difference in performance between US-S and Unet-S is indeed related to the usage of the F0 information. In Unet, the F0 information is globally used. In our case, the F0 information is directly exploited to produce harmonic signals using the explicit source production models. The neural network only infers the remaining information (vocal tract filter, noise content, etc). This makes the task easier for the neural network which can explain the substantially higher performance.


To sum up, the proposed unsupervised model-based deep learning approach to source separation performs better than learning-free and supervised purely data-driven baselines. It is also extremely efficient in learning from data. The method is useful in many scenarios where homogeneous sources need to be separated and/or only a very small amount of data (possibly without ground truth) is available for training. 
Besides choir separation as in our experiments, such scenarios may be the separation of lead from background vocals or of traditional music with less common instrumentation. Since only mixtures are needed for training, the proposed model may also be trained directly on the mixtures at hand which are to be separated. Given sufficient computational resources, parameter optimization may also be done directly on each test mixture individually, which would make the method learning-free.






\subsection{Robustness to F0 estimation errors}
\label{subsec:f0_robustness}

We propose herein two experiments to analyse the impact of multi-F0 estimation errors on the performances of our models, for the more complex case of 4 sources.



\subsubsection{Building a reference multi-F0 annotation}

For this analysis study, we  built a reference multi-F0 annotation for each time frame (16 ms) of the test set. 
Since we have access to the individual solo music tracks for each mixture signal, we opted for an automatic annotation of the solo tracks using a state-of-the-art pitch estimator (CREPE) \cite{kim2018crepe}. 
Our proposed models are named \textit{oracle models} when they are run using this reference multi-F0 annotation.

\subsubsection{Experiment 1: Analysis of F0 estimation errors}
This experiment aims at analysing further the evaluation results obtained in Section \ref{subsec:exp_results} for the model US-F. 
Using the reference multi-F0 annotation described above as ground truth, we can identify the errors made by our multi-F0 estimator \cite{cuesta2020multiple} in our experiments for each time frame of 16 ms. A \textit{Voice Activity Detection (VAD) error} is observed where one of the sources is declared active in the reference multi-F0 annotation and not by our multi-F0 estimator or vice-versa.  When there are no VAD errors, it is possible to evaluate the \textit{F0-precision} which is defined as the absolute deviation in Hertz between the value given by the reference multi-F0 annotation and our multi-F0 estimator. 

\begin{figure}
  \centering
   \includegraphics[width=\linewidth]{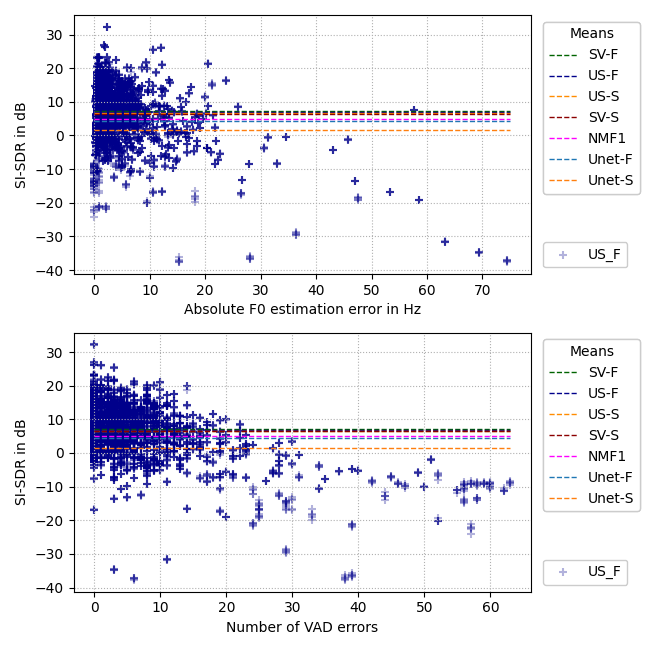}
 \caption{Separation performance for the US-F model for each evaluation frame of 1 second of the test set (4 sources)
as a function of F0 precision (top) and VAD errors (bottom).}
  \label{fig:VAD}
\end{figure}

\begin{figure}
  \centering
   \includegraphics[width=\linewidth]{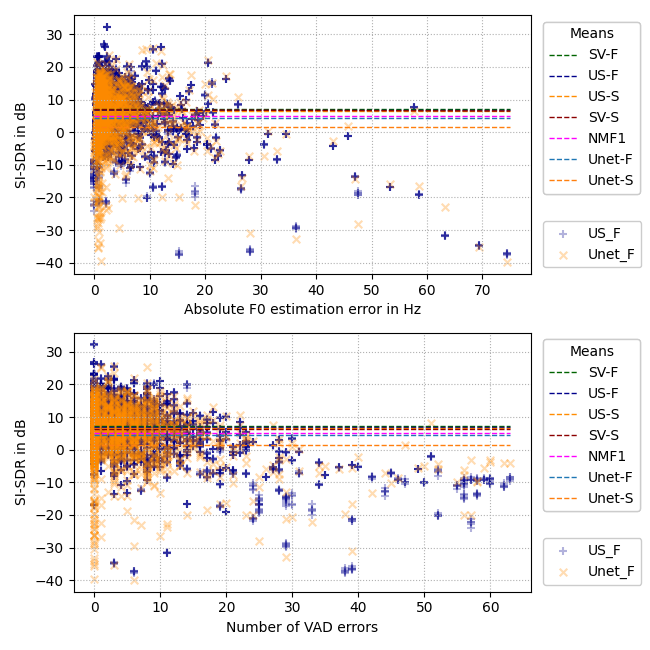}
 \caption{Comparison of the separation performances for the US-F and Unet-F models for each evaluation frame of 1 second of the test set (4 sources) as a function of F0 precision (top) and VAD errors (bottom).}
  \label{fig:VAD2}
\end{figure}

Each evaluation frame of 1 second of the test set can then be labelled with the number of effective VAD errors and the mean F0-precision for the correctly detected voices. The computed SI-SDR obtained by our model US-F for each of these frames are displayed on Fig. \ref{fig:VAD}. It can be observed that the performances are, as expected, impacted by F0 estimation errors. The degradation of performances remains limited for frames with moderate amount of errors and the algorithm is more robust to VAD errors. 
\color{black}
The performances distribution for the supervised methods are similar overall but, in the case of the Unet methods, there are more frames with low SI-SDR when there are no VAD/F0 errors (see Fig. \ref{fig:VAD2}). The  unsupervised approaches seem to benefit more than the supervised approaches from a correct estimation of VAD and F0, as already discussed in section \ref{subsec:exp_results}, but they are slightly more fragile in the case of severe estimation errors.
\color{black}

Note that the level of SI-SDR$=-7.23$dB corresponds to the result obtained on the test set by a "dummy" separator where all estimated sources are attenuated replicas of the mixture. It can then be noticed that our model is inoperative or detrimental (SI-SDR $\leq -7.23 $dB) only on a very limited number of evaluation frames, which mostly correspond to frames with particularly high numbers of VAD errors or high F0 estimation deviations. 


\subsubsection{Experiment 2: Robustness to noisy multi-F0 estimations}
We evaluate the impact of noisy F0 estimations on the performance of our models by manually degrading the reference F0 annotations.
Two different alterations are considered, namely:
\begin{itemize}
    \item \textit{Transposition:} All reference F0 frequencies are shifted upwards or downwards by a predefined number of Hertz or by one or several octaves
    \item \textit{Voice-muting:} All voices for a given time frame of 16 ms have a given probability to be muted (the corresponding multi-F0 values are set to zero) forcing as a consequence a predefined VAD error rate. 
\end{itemize}

\begin{figure}
  \centering
   \includegraphics[width=\linewidth]{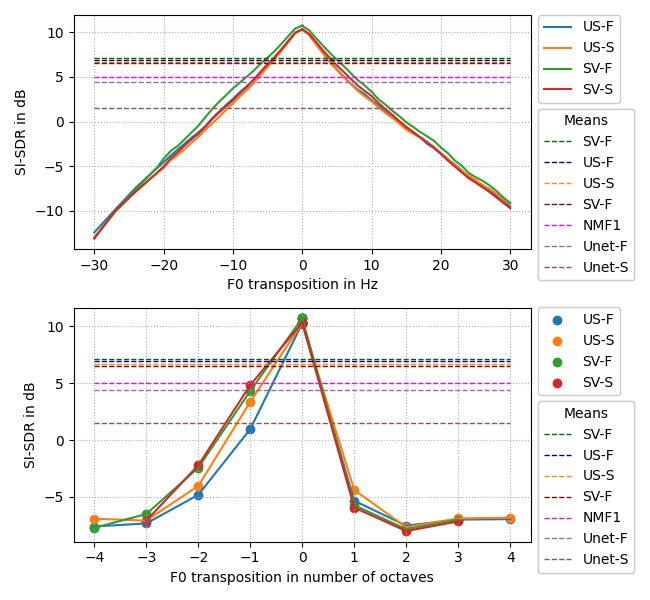}
 \caption{Effect of F0 values transposition. (top) transposition in Hz (bottom) transposition in octaves. The dotted lines show the mean SI-SDR using the multi-F0 estimation method without manual degradation (cf. Fig. \ref{fig:si_sdr})}
  \label{fig:Transposed}
\end{figure}

The results obtained for the effect of transposition are given in Fig. \ref{fig:Transposed}.  
First, it can be seen that the oracle models (our proposed models using the reference F0 values without degradation) obtain better results than the same models using F0 values given by our multi-F0 estimator (more than +3 dB on average). Second, the degradation of the performance is smoothly varying with the precision in Hz of the F0 values demonstrating the degree of tolerance of the algorithm with respect to multi-F0 estimation errors.  Third, for a perturbation of less than 5 Hz, the model is still outperforming all baseline models. And finally, the models do not increase the mean SI-SDR ($= -7.23 $dB) of the "dummy" separator 
only when the  shift exceeds roughly 23 Hz (which corresponds to at least a semi-tone for all notes below A4 (440Hz) which can be considered a severe error).
For octave errors, the degradation is very rapid when the F0 values are transposed upwards but remains moderate when transposing downwards. This may be explained by the fact that when the fundamental frequency is twice as low as the true value, every even harmonic of the source falls exactly at the position of the true target source. 

\begin{figure}
  \centering
  \includegraphics[width=\linewidth]{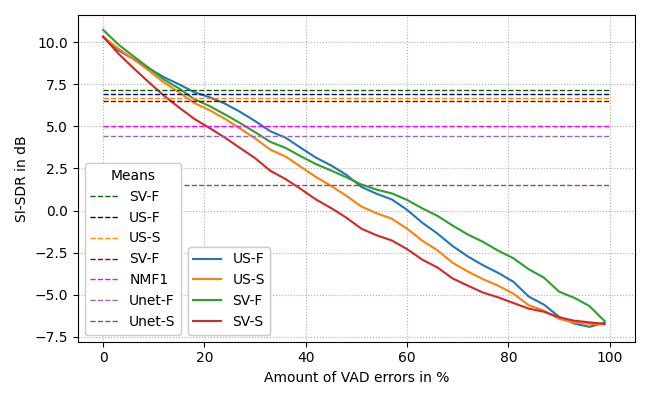}
 \caption{Effect of voice muting}
  \label{fig:muted}
\end{figure}

Fig. \ref{fig:muted} gives the results when one or several voices are locally muted. The models remain efficient when the percentage of VAD errors is below 20\%. The roughly linear shape of the degradation curve underlines the robustness of the model to such detection errors. The models trained on the full training set (US-F and SV-F) are more robust, but the supervised model seems more fragile than the unsupervised model when trained with the small dataset.

\subsection{Limitations and perspectives}
The experimental evaluation showed many advantages of the proposed approach compared to various alternatives. Nevertheless, there are some limitations. 
\color{black}{First, our approaches assume that the number of sources is known. Although we have shown that our methods are somewhat robust to moderate VAD errors, they are currently limited in cases where the number of sources is unknown. However, the most striking limitation is more precisely} \color{black}
that the method requires F0 estimates which are assigned to the sources. As for all F0-informed separation methods, the sources should exhibit mainly harmonic content and be monophonic so that the separation can be guided by the F0 information. It requires that good F0 estimates can be obtained for all sources from the mixture. As shown in the experiments, this is possible with existing methods. Progress in research on multiple F0 estimation may lead to further improvements. An extension of our method to polyphonic sources as well as estimating the F0 jointly with the other source parameters may be an interesting direction for future work. 
\color{black}In its current form, the model would not perform well on more diverse mixtures of music sources as are for example contained in the popular MUSDB dataset \cite{musdb18} which includes drums, inharmonic and polyphonic sources. \color{black} 
Moreover, audio effects such as reverberation or distortion, which may have been applied to the sources, should be explicitly modeled in the source models and must hence be known beforehand. 
Lastly, the space complexity grows linearly with the number of sources to be modeled.

In the experiments above, the final source estimates were obtained by Wiener filtering of the mixture. To this end, soft masks were obtained from the source signals $\tilde{s}_j$ generated by the source models. We also evaluated the quality of the generated signals $\tilde{s}_j$ as source estimates. The metric used for this evaluation was the spectral source-to-noise ratio \cite{virtanen2007monaural}. It can be seen as a SI-SDR which is computed on magnitude spectrograms. We used this spectral metric because the phase of the generated signals is known not to be the same as the one of the ground truth signals. This makes a time-domain evaluation not applicable.

In terms of this metric, the quality of such source estimates was inferior to the baselines and to $\hat{s}_j$ obtained using soft masks. This is because the synthesis of the signals $\tilde{s}_j$ is less constrained than masking of the mixture. The output of masking is limited by the frequency content of the mixture, since masking can only keep or remove (but not add) such content. In contrast, frequency content which is not present in any source can be contained in $\tilde{s}_j$. In fact, the DNN tends to overestimate the noise content of the sources. While this is clearly audible in $\tilde{s}_j$, no noise is added in $\hat{s}_j$.

Nevertheless, we believe that source estimates generated by parametric models are a worthwhile goal for future research. They provide a complete parameterization of the mixture signal which can be exploited for tasks such as timbre or style transfer, transposition, and melody editing of single sources. We included the generated source signals $\tilde{s}_j$ and their sum $\tilde{m}$ in the audio examples\footnote{\url{https://schufo.github.io/umss/}}. Moreover, we provide two examples of melody editing for which the mixture parametrization was exploited.



\section{Conclusion}
\label{sec:conclusion}
In this work, we presented a method for (musical) audio source separation which overcomes two limitations of state-of-the-art supervised deep learning methods: They do not separate homogeneous sources and require large datasets of mixtures with the corresponding sources in isolation for training. We proposed a novel unsupervised model-based deep learning approach. It integrates domain knowledge in the form of differentiable parametric source models in a data-driven method and exploits F0 information. Experiments show that it outperforms learning-free and supervised baselines. Furthermore, the method performs well even when trained on less than three minutes of audio data. It allows to apply powerful deep learning based separation in domains where training data is expensive or nonexistent.

\section*{Acknowledgment}

The authors would like to thank Emmanouil Benetos for providing the Bach Chorals and Barbershop Quartet dataset.

\ifCLASSOPTIONcaptionsoff
  \newpage
\fi



\bibliographystyle{IEEEtran}
\bibliography{bibliography}
\end{document}